\documentclass{ws-ijmpcs}

\begin{document}

\markboth{Bo-Qiang Ma} {New Chance for Researches on Lorentz
Violation}

%
\catchline{}{}{}{}{}
%

\title{NEW CHANCE FOR RESEARCHES ON LORENTZ VIOLATION}

\author{BO-QIANG MA}

\address{School of Physics and State Key Laboratory of Nuclear Physics and
Technology, Peking University, Beijing 100871, China\\ Center for
High Energy Physics, Peking University, Beijing 100871, China \\
mabq@pku.edu.cn}

\maketitle

\begin{history}
\received{29 February 2012}
\end{history}

\begin{abstract}
I present a brief review on the motivation for the study on Lorentz
violation and on some of our studies with phenomenological analysis
of Lorentz violation effects. I also discuss three effective field
theory frameworks for Lorentz violation: the Coleman-Glashow model,
the standard model extension (SME), and the standard model
supplement (SMS). The situation of the OPERA ``anomaly" is also
briefly reviewed, together with some discussion on the
superluminality of neutrinos within the effective field theory
frameworks.

\keywords{Lorentz violation; effective field theory; superluminal
neutrino}
\end{abstract}

\ccode{PACS Nos.: 11.30.Cp, 12.60.-i, 14.60.Lm, 14.60.St}

\section{Motivation for Lorentz Violation Study}

In 1905, Einstein published his famous paper ``On the
Electrodynamics of Moving Bodies" and established the theory of
special relativity. The special relativity offers a revolution to
the concepts of space and time in Newton's mechanics, and provides a
proof to adopt the Lorentz transformation for the covariance of the
equations of electrodynamics, to replace the traditional Galileo
transformation in classical physics. There are two basic principles
of special relativity:
\begin{itemize}
\item
 Principle of
Relativity: the equations describing the laws of physics have the
same form in all admissible frames of reference.
\item
Principle of constant  light speed: the speed of light is the same
in all directions in vacuum in all reference frames, regardless
whether the source of the light is moving or not.
\end{itemize}
Through over one hundred years of investigations from various
aspects, Einstein's relativity has won great triumphs. It becomes
one of the foundations of modern physics and has been proved to be
valid at very high precision. Lorentz Invariance, the basic
theoretical foundation of relativity, states that the equations
describing the laws of physics have the same form in all admissible
reference frames, or physical laws keep invariant under the Lorentz
transformation. So we need to answer the question: why we seek for
Lorentz violation?

The Lorentz symmetry is a symmetry related with space and time,
therefore the Lorentz violation should be related to the basic
understandings of space and time. From the viewpoint of physics, the
origin for the breaking down of conventional concepts of space and
time might be traced back to Planck. There are four basic constants
before Planck's creation of his quantum theory of black body in
1990: the Newton gravitational constant $G$,  the light speed in
vacuum $c$, the Boltzmann constant $k_B$, and the permittivity of
free space $\epsilon_0$.  In 1899, Planck introduced a new constant
$\hbar$, for the purpose to construct a ``God-given'' unit
system\cite{p99}. Then he set the above five constants as bases, and
elegantly simplified recurring algebraic expressions in physics. One
year later Planck found that the constant $\hbar$ he introduced for
his unit system is an indispensable constant for his new quantum
theory. There are a number of basic quantities in this unit system,
such as the Planck length $l_{\rm P} \equiv \sqrt{G\hbar/c^3} \simeq
1.6 \times 10^{-35}$~m, the Planck time $t_{\rm P} \equiv
\sqrt{G\hbar/c^5} \simeq 5.4 \times 10^{-44}$~s, the Planck energy
$E_{\rm P} \equiv \sqrt{\hbar c^5/G} \simeq 2.0 \times 10^{9}$~J,
and the Planck temperature $T_{\rm P} \equiv \sqrt{\hbar c^5/Gk_B^2}
\simeq 1.4 \times 10^{32}$~K. Therefore one may suspect that
conventional understanding of space and time might be breaking down
at the Planck scale\cite{lv5}: i.e., at the Planck length $l_{\rm
P}$, or the Planck time $t_{\rm P}$, or the Planck energy $E_{\rm
P}$, where new features of existence may emerge. The breaking down
of continue space-time was also conjectured\cite{Snyder,Wheeler}.

Just recently, Xu and I provided a physical argument for the
discreteness of space and time\cite{xu-l}. From two known entropy
constraints:
\begin{equation}
S_{\mathrm{matter}} \leq 2\pi E R, ~~~~  \mathrm{and}
~~S_{\mathrm{matter}} \leq \frac{A}{4},
\end{equation}
combined with the black-body entropy,
\begin{equation} S =
\frac{4}{45}{\pi}^2 {T}^3 V = \frac{16}{135}{\pi}^3 R^3 T^3,
\end{equation}
we arrive at a minimum value of space
\begin{equation}
 R\geq \Big(\frac{128}{3645\pi}\Big)^{\frac{1}{2}} l_{\mathrm{P}} \simeq 0.1
 l_{\mathrm{P}}.
 \end{equation}
Thus we reveal from physical arguments that space-time is discrete
rather than continuous. From another point of view, the newly
proposed entropic gravity suggests gravity as an emergent force
rather than a fundamental one\cite{Verlinde,HeMa}. If gravity is
emergent, a new fundamental constant should be introduced to replace
the Newtonian constant $G$\cite{lv5}. It is natural to suggest a
fundamental length scale, and such constant can be explained as the
smallest length scale of quantum space-time. Its value can be
measured through searches of Lorentz violation\cite{lv5,xu-l}. The
existence of an ``{\ae}ther" or ``vacuum" can also bring the
breaking down of Lorentz invariance\cite{Dirac,Bjorken}.

Thus the research on the Lorentz violation may provide us the chance
for new understanding of the nature of basic concepts, such as
``space", ``time", and ``vacuum", through physical ways, rather than
from the viewpoint of metaphysics or philosophy.

Nowadays, there has been an increasing interest in Lorentz
invariance Violation (LV or LIV) both theoretically and
experimentally. The possible Lorentz symmetry violation (LV) effects
have been sought for decades from various theories, motivated by the
unknown underlying theory of quantum gravity together with various
phenomenological
applications\cite{ShaoMa10,lv3,Shao2010,Shao2011,lv4,Bietenholz:2008ni}.
This can happen in many alternative theories, e.g., the doubly
special relativity
(DSR)\cite{Amelino-Camelia2002,Magueijo:2001cr,Zhang2011}, effects
in general relativity\cite{LV-GR1,Ni:2009fg,LV-GR2}, non-covariant
field
theories\cite{Copenhagen1,Copenhagen2,Copenhagen3,Copenhagen4}, and
large extra-dimensions\cite{Ammosov2000,Pas2005}. As examples, I
list below some phenomenological consequences of the Lorentz
violation effects studied by my students and I in the last a few
years:
\begin{itemize}
\item
The Lorentz violation could provide an explanation of neutrino
oscillation without neutrino mass~\cite{Xiao08,Yang09}. We carried
out Lorentz violation contribution to neutrino oscillation by the
effective field theory for Lorentz violation and give out the
equations of neutrino oscillation probabilities. In our model,
neutrino oscillations do not have drastic oscillation at low energy
and oscillations still exist at high energy. It is possible that
neutrinos may have small mass and both Lorentz violation and the
conventional oscillation mechanisms contribute to neutrino
oscillation.

\item
The modified dispersion relation of the proton could increase the
threshold energy of photo-induced meson production of the proton and
cause an increase of the GZK cutoff energy. The earlier reports on
super-GZK events triggered attention on Lorentz-Violation. The new
results of observation of GZK cut-off put strong constraints on
Lorentz violation parameters\cite{Xiao08}.
\item
The modified dispersion relation of the photon may cause time lag of
photons with different energies when they propagate in space from
far-away astro-objects. The Lorentz violation can modify the photon
dispersion relation, and consequently the speed of light becomes
energy-dependent\cite{lv3}. This results in a tiny time delay
between high energy photons and low energy ones. Very high energy
photon emissions from cosmological distance can amplify these tiny
LV effects into observable quantities. We analyzed photons from
$\gamma$-ray bursts from Fermi satellite observations and presented
a first robust analysis of these taking the intrinsic time lag
caused by sources into account, and gave an estimate to LV energy
scale $\sim 2 \times 10^{17}$~GeV for linear energy dependence, and
$\sim 5 \times 10^9$~GeV for quadratic dependence\cite{Shao2010}.
\item
We also studied recent data on Lorentz violation induced vacuum
birefringence from astrophysical consequences\cite{Shao2011}. Due to
the Lorentz violation, two helicities of a photon have different
phase velocities and group velocities, termed as ``vacuum
birefringence''. From recently observed $\gamma$-ray polarization
from Cygnus X-1, we obtained an upper limit $\sim 8.7\times10^{-12}$
for Lorentz-violating parameter $\chi$, which is the most firm
constraint from well-known systems.
\end{itemize}

\section{Lorentz Violation in Effective Field Theory Frameworks}

Among many theoretical investigations of Lorentz violation, it is a
powerful framework to discuss various LV effects based on
traditional techniques of effective field theory in particle
physics. The general effective field theory framework starts from
the Lagrangian of the standard model, and then includes all possible
terms containing the Lorentz violation effects. The magnitudes of
these LV terms can be constrained by various experiments. In the
following we briefly discuss three different versions of effective
field theory frameworks for Lorentz violation: the Coleman-Glashow
model\cite{Coleman99}; the minimal standard model extension (SME)
\cite{Colladay1998}; and the newly proposed standard model
supplement (SMS)\cite{Ma10,SMS3}.

\subsection{The Coleman-Glashow model}

The Coleman-Glashow model\cite{Coleman99} is a simple version to
include Lorentz violating terms into the standard model Lagrangian.
Let $\Psi$ denote a set of $n$ complex scalar fields assembled into
a column vector, one can add to the standard model Lagrangian the
Lorentz-violating term:
\begin{equation}
{\cal L}\rightarrow{\cal L} +
\partial_i\Psi \epsilon\partial^i\Psi,
\label{Coleman-Glashow}
\end{equation}
where $\epsilon$ is a Hermitian matrix that signals the Lorentz
violation in the Coleman-Glashow model. In case for fermions, the
wave function $\Psi$ denotes the Dirac spinor, and the LV parameter
$\epsilon$ can be taken as a fixed scaler constant for the fermion
under consideration.

We should stress that the Coleman-Glashow model can be only
considered as a ``toy model" for illustration as it does not meet
the criterion of being an ``exact" theory. The reason is that the
simple form of the Lagrangian Eq.~(\ref{Coleman-Glashow}) can not be
taken as invariant in all inertial frames of reference, but only
valid in one inertial frame of reference the observer is working.
This model can be adopted when the observer is focusing on the
Lorentz violation effect within a certain frame such as the
earth-rest frame, the sun-rest frame, or the CMB frame, and does not
care about relationships between different frames. Otherwise the
situation could become very complicated with different formalisms in
different frames from the requirement of consistency, i.e., the
absolute physical events should keep unchanged no matter observed
from any reference frame. We can consider this requirement as a
basic principle called absolute physical event consistency:
\begin{itemize}
\item
For a physical process when the initial and final particles are
experimentally produced and detected, such event is called an
absolute physical event, and its existence does not change when
viewed from another frame of reference.
\end{itemize}
For example, the decay of a physical particle, such as $\pi \to \mu
+\nu_{\mu}$, if happens in one frame from an observer, it should
also happen observed from another frame by another observer.
However, the situation might be different for virtual processes. The
virtual processes could be different when viewed from different
frames.

Physically, the reason for the Lorentz violation of the
Coleman-Glashow model is due to the existence of a ``background"
represented by the scalar constant $\epsilon$. This model can apply
as a practical tool to demonstrate or estimate the magnitude of the
Lorentz violation effect. It has been successfully applied in many
phenomenological analysis to constrain the Lorentz violation effect
in some physical processes.

\subsection{The minimal standard model extension (SME)}

In the standard model extension (SME), terms that violate Lorentz
invariance can be added by hands, and then one can select the terms
from some considerations such as gauge invariance, Hermitean,
power-counting renormalizability and etc. In the minimal version of
the SME\cite{Colladay1998}, the LV terms are measured with several
tensor fields as coupling constants, and modern experiments have
built severe constraints on the relevant Lorentz violation
parameters\cite{Kostelecky2011}.

For example, the SME Lagrangian for massless neutrinos takes the
form\cite{Colladay1998,Yang09,Qin-OPERA}
\begin{align}
\mathcal{L}=\frac{1}{2}i\overline{\nu}_{A}\gamma^{\mu}\overleftrightarrow{D_{\mu}}\nu_{_B}\delta_{_{AB}}
+\frac{1}{2}ic_{_{AB}}^{\mu\nu}\overline{\nu}_{_A}\gamma_{\mu}\overleftrightarrow{D_{\nu}}\nu_{_B}
-a_{_{AB}}^{\mu}\overline{\nu}_{_A}\gamma_{\mu}\nu_{_B}+\cdots\;,\label{Lagrangian}
\end{align}
where $c_{AB}^{\mu\nu}$ and $a_{AB}^{\mu}$ are Lorentz violation
coefficients which can be thought as resulting from tensor vacuum
expectation values in some kind of underlying theory, the $A,B$ are
flavor indices, and the ellipsis denotes the non-renormalizable
operators (eliminated in the minimal SME). The first term in
Eq.~(\ref{Lagrangian}) is exactly the SM operator, and the second
and third terms (CPT-even and CPT-odd respectively) describe the
contribution from Lorentz violation. The coefficients
$c_{AB}^{\mu\nu}$ and $a_{AB}^{\mu}$ serve as the background fields
though they are not fields but general tensers with constant values
in the observer working frame.

In SME, the background fields transform as tensors according to
their Lorentz indices between different inertial frames of reference
but keep unchanged within the same frame. It means that there exists
a privileged inertial frame of reference in which the background can
be considered as the ``new {\ae}ther", i.e., the ``vacuum" at rest.
The ``{\ae}ther", which is a collections of background fields,
changes from one frame to another frame by Lorentz transformation.
Within a same frame of reference, these background fields are just
treated as fixed parameters. The Lorentz violation is due to the
existence of the background fields. The standard model particles
breaks the Lorentz invariance at a certain frame of the observer by
taking these background fields as fixed. From a strict sense, there
is no Lorentz violation for the whole system of the standard model
particles together with the background fields.

\subsection{The standard model supplement (SMS)}

The standard model extension is an effective framework for
phenomenological analysis. We still need a fundamental theory to
derive the Lorentz violation terms from basic principles. In the
Standard Model Supplement (SMS) framework\cite{Ma10,SMS3}, the LV
terms are brought about from a basic principle denoted as the
physical independence or physical invariance:
\begin{itemize}
\item
Principle of Physical Invariance: the equations describing the laws
of physics have the same form in {\bf all admissible mathematical
manifolds}.
\end{itemize}
The principle leads to the following replacement of the ordinary
partial $\partial_{\alpha}$ and the covariant derivative
$D_{\alpha}$
\begin{equation}\label{eqn:substitution}
\partial^{\alpha} \rightarrow M^{\alpha\beta}\partial_{\beta},\quad
D^{\alpha}\rightarrow M^{\alpha\beta}D_{\beta},
\end{equation}
where $M^{\alpha\beta}$ is a local matrix. The Lorentz violation
terms are thus uniquely determined from the standard model
Lagrangian without any ambiguity\cite{Ma10}, and their general
existence is derived from basic consideration rather than added by
hand. The explicit form of the matrices $M^{\alpha\beta}$ demands
more basic theories concerning the true nature of space and time,
and we suggest to adopt a physical way to explore these matrices
through experiments rather than from theory at first. For more
generality, we do not make any ad hoc assumption about these
matrices. Thus these matrices might be particle dependent
corresponding to the standard model particles under consideration,
with the elements of these matrices to be measured or constrained
from experimental observations.

We separate $M^{\alpha\beta}$ to two matrices like $M^{\alpha
\beta}=g^{\alpha \beta}+\Delta^{\alpha \beta}$, where
$g^{\alpha\beta}$ is the metric tensor of space-time and
$\Delta^{\alpha \beta}$ is a new matrix which is particle-type
dependent generally. Since $g^{\alpha\beta}$ is Lorentz invariant,
$\Delta^{\alpha \beta}$ contains all the Lorentz violating degrees
of freedom from $M^{\alpha \beta}$. Then $\Delta^{\alpha \beta}$
brings new terms violating Lorentz invariance in the standard model
and is called Lorentz violation matrix. The theory returns back to
the standard model when these Lorentz violation matrices vanish.
Thus one may consider the Lorentz violation matrices in the SMS
framework as similar to the background fields in the minimal SME
model.

For the electroweak interaction sector, the Lagrangian of fermions
in the SMS framework can be written as\cite{Ma10,SMS3,SMS-OPERA}
\begin{eqnarray}\label{Lagrangian_F}
\mathcal{L}_{\mathrm{F}} &=&
i\bar{\psi}_{A,\mathrm{L}}\gamma^{\alpha}\partial_{\alpha}\psi_{B,\mathrm{L}}\delta_{AB}+
i\Delta^{\alpha\beta}_{\mathrm{L},AB}\bar{\psi}_{A,\mathrm{L}}\gamma_{\alpha}\partial_{\beta}\psi_{B,\mathrm{L}}\nonumber\\
&&
+i\bar{\psi}_{A,\mathrm{R}}\gamma^{\alpha}\partial_{\alpha}\psi_{B,\mathrm{R}}\delta_{AB}+
i\Delta^{\alpha\beta}_{\mathrm{R},AB}\bar{\psi}_{A,\mathrm{R}}\gamma_{\alpha}\partial_{\beta}\psi_{B,\mathrm{R}},
~~~~
\end{eqnarray}
where $A,B$ are flavor indices. The Lorentz violation terms are
uniquely and consistently determined from the standard model by
including the Lorentz violation matrices $\Delta^{\alpha\beta}$,
which are generally particle-dependent\cite{SMS3} with flavor
indices. For leptons, $\psi_{A,\mathrm{L}}$ is a weak isodoublet,
and $\psi_{A,\mathrm{R}}$ is a weak isosinglet. After the
calculation of the doublets and classification of the Lagrangian
terms again, the Lagrangian can be written in a form like that of
Eq.~(\ref{Lagrangian_F}) too. Assume that the Lorentz violation
matrix $\Delta^{\alpha\beta}_{AB}$ is the same for the
left-handedness and right-handedness, namely
$\Delta^{\alpha\beta}_{\mathrm{L},AB}=\Delta^{\alpha\beta}_{\mathrm{R},AB}=\Delta^{\alpha\beta}_{AB}$.
Without considering mixing between flavors, one can rewrite
Eq.~(\ref{Lagrangian_F}) as
\begin{equation}\label{Lagrangian_F_total}
\mathcal{L}_{\mathrm{F}}=\bar{\psi}_A
(i\gamma^{\alpha}\partial_{\alpha}-m_A)\psi_A
+i\Delta^{\alpha\beta}_{AA}\bar{\psi}_A\gamma_{\alpha}\partial_{\beta}\psi_A,
\end{equation}
where $\psi_A=\psi_{A,\mathrm{L}}+\psi_{A,\mathrm{R}}$, i.e., the
field $\psi_A$ is the total effects of left-handed and right-handed
fermions of the given flavor $A$.  When there is only one handedness
for fermions, $\psi_A$ is just the contributions of this one
handedness, which is the situation for neutrinos. The mass term in
the Lagrangian $\mathcal{L}_{\mathrm{F}}$ is included, one can let
$m_A\rightarrow 0$ for massless fermions.

In similar to the above two frameworks, there also exists the
question of how to understand and handle the Lorentz violation
matrix $\Delta^{\alpha\beta}$. We list here three options for
understandings and treatments\cite{Ma:2011jj}:

\begin{itemize}
\item
{\bf Scenario I}: which can be called as fixed scenario in which the
Lorentz violation matrices are taken as constant matrices in any
inertial frame of reference the observer is working. It means that
the the Lorentz violation matrices can be taken as approximately the
same for any working reference frames such as the earth-rest frame,
the sun-rest frame, or the CMB frame. This scenario can be only
adopted as an approximation in similar to the Coleman-Glashow model,
when the observer is focusing on the Lorentz violation effect within
a certain frame and does not care about relationships between
different frames. There will be the problem of inconsistency for an
``absolute physical event" between different reference frames as
pointed out for the Coleman-Glashow model, if one sticks to this
scenario.


\item
{\bf Scenario II}: which can be called as ``new {\ae}ther" scenario
in which the Lorentz violation matrices transform as tensors between
different inertial frames but keep as constant matrices within the
same frame. This scenario corresponds to the same treatment of
background fields as in the SME case. The Lorentz violation matrices
play the roles as the background fields.

\item
{\bf Scenario III}: which can be called as covariant scenario in
which the Lorentz violation matrices transform as tensors adhered
with the corresponding standard model particles. It means that these
Lorentz violation matrices are emergent and covariant with their
standard model particles. Such a scenario still needs to be checked
for consistency and for applications.
\end{itemize}

Before accepting the SMS as a fundamental theory, one can take the
SMS as an effective framework for phenomenological applications by
confronting with various experiments to determine and/or constrain
the Lorentz violation matrix $\Delta^{\alpha \beta}$ for various
particles. So our idea is to reveal the real structure of Lorentz
violation of nature from experiments rather than from theory. We
consider this phenomenological way as more appropriate for physical
investigations, rather than to derive everything from theory at
first. As a comparison, the specific form of quark mixing matrix is
determined from experimental measurements rather than derived from
theory\cite{Ma:2011gh}. Even after so many years of research and
also the elements of CKM mixing matrix have been measured to very
high precision, there is still no a commonly accepted theory to
derive the quark mixing matrix from basic principles.

\subsection{Some remarks}

In the effective field theory frameworks, the standard particles
transform according to the Lorentz symmetry between different
momentum states. The background fields should also transform
according to the Lorentz symmetry between different observer working
frames from the requirement of consistency. From this sense, there
is actually no Lorentz violation for the whole system of the
standard model particles together with the background fields. The
Lorentz violation exists for the standard model particles within an
observer working frame, when these particles have different momenta
between each other. From this sense, the Lorentz violation is due to
the existence of the background fields, which are treated as fixed
parameters in the observer working frame.

The above discussed three frameworks have their own advantages and
disadvantages in formalisms and in phenomenological applications.
The Coleman-Glashow model is the most simple and intuitive for fast
applications to physical processes, for the estimation of the
magnitude of the Lorentz violation effect. The SME is systematic
with all possible terms that can serve as a useful tool to confront
with various phenomenological constraints. The SMS is theory based
with clear relationship between some general LV parameters in
SME\cite{SMS-photon2}, and can be conveniently applied for
phenomenological analysis\cite{Ma10graal}. We would need more
experimental investigations to check which one of them can meet the
criterion of being able to provide a satisfactory description of the
physical reality, with simplicity and beauty in formalism, together
with the predictive power towards new knowledge for human being. It
is also possible that the nature satisfies the Lorentz symmetry
perfectly and we would be unable to find a physical evidence to
support any of these theories.

\section{The OPERA ``Anomaly"}

The report by the OPERA Collaboration for a faster-than-light speed
of muon neutrinos has attracted the eyes by the physical society as
well as the public society\cite{Ma:2011jj}. We have witnessed three
stages of the OPERA performance:
\begin{itemize}
\item
The release of the first version of the OPERA report\cite{Adam2011}
on September 22 of 2011, reporting that the muon neutrinos in the
CERN-CNGS neutrino beam were detected by the OPERA detector over a
baseline of about 730~km. Compared to the time taken for neutrinos
traveling at the speed of light in vacuum, an earlier arrival time
of $(60.7\pm6.9~(\mathrm{stat.})\pm
7.4~(\mathrm{sys.}))~{\mathrm{ns}}$ was measured. The neutrino
velocity $v$ is thus measured and its difference with respect to the
vacuum light speed $c$ is ${(v-c)}/{c}=\left(2.48 \pm 0.28
(\mathrm{stat.}) \pm 0.30 (\mathrm{sys.})\right)\times 10^{-5}$ at a
significance of $6\sigma$.
\item
To overcome the criticism that the long proton beam duration at CERN
may introduce bias in the neutrino arrival time measurement, the
OPERA collaboration released their revised report\cite{re-OPERA} on
November 17 of 2011. They repeated the measurement over the same
baseline without any assumptions about the details of neutrino
production during the spill, such as energy distribution or
production rate, by using a new CERN beam which provided proton
pulses of 3~nanoseconds each with 524~nanosecond gaps. Without using
the earlier statistical computation, the OPERA collaboration
measured twenty events indicating neutrinos had traveled faster than
light by 60~ns, with 10~ns uncertainty. The error bounds for the
original superluminal speed fraction were tightened further to
$(2.37 \pm 0.32 (\mathrm{stat.}) + 0.34/ - 0.24
(\mathrm{sys.}))¡Á10^{-5}$, with the new significance level becoming
$6.2\sigma$.
\item
There was a news on February 22 of 2012 that the OPERA collaboration
has identified two possible effects that could have an influence on
its neutrino timing measurement. The first possible effect concerns
an oscillator used to provide the time stamps for GPS
synchronizations, and the second concerns the optical fibre
connector that brings the external GPS signal to the OPERA master
clock. The two effects could have led to an underestimate of the
flight time of the neutrinos, and a re-measurement of the neutrino
speed by the OPERA collaboration will be done in the near future.
\end{itemize}

Before the OPERA ``anomaly", there have been similar long baseline
experiments on the speed of neutrinos. The first direct measurement
of neutrino velocity was performed at Fermilab thirty years
ago\cite{Alspector1976,Kalbfleisch1979}. Just a few years ago, the
MINOS Collaboration\cite{Adamson2007} reported a shift with respect
to the expected time of flight of
$\delta_t=-126\pm32~(\mathrm{stat})\pm64~(\mathrm{sys)}~\mathrm{ns}$,
which corresponds to a constraint on the muon neutrino velocity,
$(v_{\nu}-c)/c=(5.1\pm2.9)\times10^{-5}$
at $68\%$ confidence level. This $1.8\sigma$ signal was considered
to be compatible with also zero, therefore the previous experimental
data neither provide a strong evidence for the superluminality of
neutrinos nor exclude it.

Besides the long baseline experiments, there are also measurable
phenomenologies related to neutrino speed in astrophysics. For
instance, one astronomical event was observed with neutrino
emissions on 23 February 1987, 7:35:35 UT ($\pm1$ min)
--- the Supernova 1987A\cite{Hirata1987,Bionta1987}, which was later optically observed on
24 February 1987. More than ten neutrinos were recorded with a
directional coincidence within the location of supernova explosion,
several hours before the optical lights were observed. Because of
weak interactions, neutrinos may leak out of the dense environment
produced by the stellar collapse before the optical depth of photons
becomes visible. Hence an early-arrival of neutrinos is expected.
The journey of propagation of photons and neutrinos are of
astrophysical distance ($\sim 51.4$~kpc), hence it provides an
opportunity to measure\cite{Longo1987} the speed of neutrinos to be
within the light speed with a precision of $\sim2\times10^{-9}$.
This also neither provides an evidence for the superluminality of
neutrinos nor excludes it.

However, as the OPERA ``anomaly" seems to support the
superluminality of neutrinos strongly, there has been a blossom of
novel theories that can produce the superluminality of
neutrinos\cite{Ma:2011jj}. In fact, the effective field theory
framework can produce modification to the standard energy-momentum
dispersion relation of a particle. One thus can calculate the
particle velocity through the new dispersion relation, in which the
LV parameters enter. The velocity of a particle could be therefore
superluminal or subluminal by adjusting the LV
parameters\cite{Ma:2011jj}. By confronting with the OPERA reported
``data", the LV parameters were estimated in Ref.~\cite{SMS-OPERA}
for the SMS framework and in Ref.~\cite{Qin-OPERA} for the minimal
SME, indicating a magnitude of the order $10^{-5}$ for relevant LV
parameters in both frameworks.

With a size of order $10^{-5}$ for the LV parameter $\epsilon$,
Cohen and Glashow argued\cite{Glashow11} that the high energy muon
neutrinos exceeding tens of GeVs can not be detected due to the
energy-losing process $\nu_\mu\rightarrow\nu_\mu+e^{+}+e^{-}$
analogous to Cherenkov radiations through the long baseline about
730~km. Bi {\it et al.} also argued that the Lorentz violation of
muon neutrinos of order $10^{-5}$ will forbid kinematically the
production process of muon neutrinos $\pi\rightarrow \mu + \nu_\mu$
for muon neutrinos with energy larger than about 5~GeV\cite{Bi11}.
Their arguments provide demonstrations of adopting the
Coleman-Glashow model for fast and intuitive illustrations of the LV
effects, and the arguments have been also adopted as a reason to
refute the OPERA ``anomaly". The conclusion of the Cherenkov-like
radiations and the forbidden of the muon neutrino production
processes can be also true in the SMS and SME frameworks with the LV
parameters fixed in the Scenarios I and II as discussed in the last
section. The rationality of superluminality of neutrinos seems to be
only possibly in the covariant picture of Lorentz violation, such as
in the Scenario III suggested as an option to handle the LV effect
in the effective field theory framework\cite{Ma:2011jj,SMS-OPERA2}.
However, such a possibility can only be seriously considered when
there will be strong evidence for the superluminality of neutrinos
in future experiments.

\section{Conclusion}

Researches on Lorentz violation have been active for many years,
with various theories have been proposed and many phenomenological
studies have been performed to confront with various observations,
though there is still no convincing evidence yet. However, there are
new chances for Lorentz violation study due the availability of many
new theoretical frameworks that can be applied to phenomenological
analysis more conveniently and also due to the developments of high
precision measurements for the future experimental investigations.
We conclude that Lorentz violation is becoming an active frontier to
explore both theoretically and experimentally.

\section*{Acknowledgements}
I am very grateful for the discussions and collaborations with a
number of my students: Zhi Xiao, Shi-Min Yang, Lijing Shao, Lingli
Zhou, Xinyu Zhang, Yunqi Xu, and Nan Qin, who devoted their wisdoms
and enthusiasms bravely on the topic of Lorentz violation. The
content of this contribution is mainly based on the collaborated
works with them. I am also very indebted to many colleagues for the
valuable discussions with them during some workshops and seminars.
The work was supported by National Natural Science Foundation of
China (Nos.~10975003, 11021092, 11035003 and 11120101004).



\begin{thebibliography}{0}    

\bibitem{p99}
M. Planck, Sitzber.\ K.\ Preuss Aka.\ Berlin {\bf 5}, 440 (1899).

\bibitem{lv5}
  L.~Shao, B.-Q.~Ma,
  Sci. China Phys. Mech. Astro. {\bf 54}, 1771 (2011)
  [arXiv:1006.3031 [hep-th]].

\bibitem{Snyder}
H.S. Snyder, Phys. Rev. 71, 38 (1947); 72, 68 (1947).

\bibitem{Wheeler}
J.A. Wheeler,
Ann. Phys. 2, 604 (1957).

\bibitem{xu-l}
  Y.~Xu and B.-Q.~Ma,
  Mod.\ Phys.\ Lett.\  A {\bf 26}, 2101 (2011)
  [arXiv:1106.1778 [hep-th]].

\bibitem{Verlinde}
  E.~P.~Verlinde,
  {\it JHEP} {\bf 1104}, 029 (2011),
  arXiv:1001.0785 [hep-th].

\bibitem{HeMa}
  X.~G.~He and B.-Q.~Ma,
   {\it Chin.\ Phys.\ Lett.\ }  {\bf 27}, 070402 (2010)
  [arXiv:1003.1625 [hep-th]].



\bibitem{Dirac}
P.A.M. Dirac,
Nature {\bf 168}, 906 (1951).

\bibitem{Bjorken}
J.D. Bjorken,
Ann.\ Phys. {\bf 24}, 174 (1963).


\bibitem{ShaoMa10}
For a brief review on Lorentz violation effects through very high
energy photons of astrophysical sources, see. e.g.,
  L.~Shao, B.-Q.~Ma,
Mod.\ Phys.\ Lett.\  A {\bf 25}, 3251 (2010) [arXiv:1007.2269], and
references therein.


\bibitem{lv3}
  Z.~Xiao, B.-Q.~Ma,
  Phys.\ Rev.\  {\bf D80}, 116005 (2009).

\bibitem{Shao2010}
  L.~Shao, Z.~Xiao, B.-Q.~Ma,
  Astropart.\ Phys.\  {\bf 33}, 312 (2010).

\bibitem{Shao2011}
  L.~Shao and B.~-Q.~Ma,
  Phys.\ Rev.\ D\ {\bf 83}, 127702  (2011)
  [arXiv:1104.4438 [astro-ph.HE]].

\bibitem{lv4}
  Z.~Xiao, L.~Shao, B.-Q.~Ma,
  Eur.\ Phys.\ J.\  {\bf C70}, 1153 (2010).





\bibitem{Bietenholz:2008ni}
  W.~Bietenholz,
  Phys.\ Rept.\  {\bf 505}, 145 (2011).



\bibitem{Amelino-Camelia2002}
  G.~Amelino-Camelia,
  Int.\ J.\ Mod.\ Phys.\  D {\bf 11}, 35 (2002)
  [arXiv:gr-qc/0012051].

\bibitem{Magueijo:2001cr}
  J.~Magueijo and L.~Smolin,
  Phys.\ Rev.\ Lett.\  {\bf 88}, 190403 (2002)
  [arXiv:hep-th/0112090].

\bibitem{Zhang2011}
  X.~Zhang, L.~Shao, B.-Q.~Ma,
  Astropart.\ Phys.\  {\bf 34}, 840 (2011).

\bibitem{LV-GR1}
W.-T. Ni, Phys.\ Rev.\ Lett.\ {\bf 38}, 301 (1977).

\bibitem{Ni:2009fg}
  W.-T. Ni,
  Rept.\ Prog.\ Phys.\  {\bf 73}, 056901 (2010)
  [arXiv:0912.5057 [gr-qc]].

\bibitem{LV-GR2}
M.L. Yan,
Commun.\ Theor.\ Phys.\ {\bf 2}, 1281 (1983).

\bibitem{Copenhagen1}
H.B. Nielsen and M. Ninomiya,  Nucl. Phys. B 141, 153 (1978).

\bibitem{Copenhagen2}
S. Chadha and H.B. Nielsen, Nucl. Phys. B 217, 125 (1983).

\bibitem{Copenhagen3}
H.B. Nielsen and I. Picek, Phys. Lett. 114B, 141 (1982).

\bibitem{Copenhagen4}
H.B. Nielsen and I. Picek, Nucl. Phys. B211, 269 (1983).

\bibitem{Ammosov2000}
V.~Ammosov and G.~Volkov, hep-ph/0008032.

\bibitem{Pas2005}
H. Pas, S. Pakvasa, T.J. Weiler,  Phys.Rev. D {\bf 72}, 095017
(2005).

\bibitem{Xiao08}
  Z.~Xiao, B.-Q.~Ma,
  Int.\ J.\ Mod.\ Phys.\  A {\bf 24}, 1359 (2009)
  [arXiv:0805.2012].

\bibitem{Yang09}
S. Yang, B.-Q.~Ma, Int. J. Mod. Phys. A {\bf24}, 5861 (2009)
[arXiv:0910.0897].


\bibitem{Coleman99}
S.R. Coleman, S.L. Glashow, Phys. Rev. D {\bf59}, 116008 (1999).


\bibitem{Colladay1998}
  D.~Colladay and V.~A.~Kostelecky,
  Phys.\ Rev.\  D {\bf 58}, 116002 (1998).



\bibitem{Ma10}
Zhou L., B.-Q. Ma, Mod. Phys. Lett. A {\bf25}, 2489 (2010)
[arXiv:1009.1331].

\bibitem{SMS3}
Zhou L., B.-Q.~Ma, Chin.\ Phys.\ C (HEP \& NP) {\bf 35}, 987 (2011)
[arXiv:1109.6387].



\bibitem{SMS-OPERA}
Zhou L., B.-Q. Ma, arXiv:1109.6097.




\bibitem{Kostelecky2011}
V. A. Kostelecky, N. Russell, Rev. Mod. Phys. {\bf83}, 11 (2011)
[arXiv:0801.0287].

\bibitem{Qin-OPERA}
  N.~Qin and B.-Q.~Ma, Int. J. Mod. Phys. A {\bf 27}, 1250045 (2012)
  [arXiv:1110.4443 [hep-ph]].


\bibitem{Ma:2011jj}
For a brief review on experiments for superluminal neutrinos and
some theoretical investigations,  see, e.g.,
B.-Q. Ma,
 {\it Mod.\ Phys.\ Lett.\  A} {\bf 27}, 1230005 (2012).


\bibitem{Ma:2011gh}
For a brief review on some standard and non-standard researches on
fermion mixing matrices, see, e.g.,
  B.-Q.~Ma,
  Int.\ J.\ Mod.\ Phys.\ Conf.\ Ser.\  {\bf 1}, 291 (2011)
  [arXiv:1109.5276 [hep-ph]].



\bibitem{SMS-photon2}
  Zhou L., B.-Q.~Ma,
  arXiv:1110.1850 [hep-ph].




\bibitem{Ma10graal}
Zhou L., B.-Q. Ma, arXiv:1009.1675.



\bibitem{Adam2011}
  T.~Adam {\it et al.}  [OPERA Collaboration],
  arXiv:1109.4897 [hep-ex].



\bibitem{re-OPERA}
Please see the second version of the OPERA paper, arXiv:1109.4897v2,
released on November 18, 2011.



\bibitem{Alspector1976}
  J.~Alspector {\it et al.},
  Phys.\ Rev.\ Lett.\ \ {\bf 36}, 837  (1976).

\bibitem{Kalbfleisch1979}
  G.~R.~Kalbfleisch, N.~Baggett, E.~C.~Fowler and J.~Alspector,
  Phys.\ Rev.\ Lett.\ \ {\bf 43}, 1361  (1979).

\bibitem{Adamson2007}
  P.~Adamson {\it et al.} [MINOS Collaboration],
  Phys.\ Rev.\ D\ {\bf 76}, 072005  (2007)
  [arXiv:0706.0437 [hep-ex]].

\bibitem{Hirata1987}
  K.~Hirata {\it et al.} [KAMIOKANDE-II Collaboration],
  Phys.\ Rev.\ Lett.\ \ {\bf 58}, 1490  (1987).

\bibitem{Bionta1987}
  R.~M.~Bionta {\it et al.},
  Phys.\ Rev.\ Lett.\ \ {\bf 58}, 1494  (1987).

\bibitem{Longo1987}
  M.~J.~Longo,
  Phys.\ Rev.\ D\ {\bf 36}, 3276  (1987).





\bibitem{Glashow11}
A.G. Cohen, S.L. Glashow, Phys. Rev. Lett. {\bf107}, 181803 (2011)
[arXiv:1109.6562].

\bibitem{Bi11}
  X.-J.~Bi, P.-F.~Yin, Z.-H.~Yu and Q.~Yuan,  Phys. Rev. Lett. {\bf 107}, 241802 (2011)
  [arXiv:1109.6667].


\bibitem{SMS-OPERA2}
Zhou L., B.-Q.~Ma, arXiv:1111.1574.


\end{thebibliography}
\end{document}